# Photoluminescence enhancement in quaternary III-nitrides alloys grown by molecular beam epitaxy with increasing Al content

S. Fernández-Garrido[a)], J. Pereiro, F. González-Posada, E. Muñoz, and E. Calleja

*ISOM and Dpt. de Ingeniería Electrónica, ETSI Telecomunicación, Universidad Politécnica de Madrid, Ciudad Universitaria s/n, 28040 Madrid Spain*

A. Redondo-Cubero, and R. Gago

*Centro de Micro-Análisis de Materiales, Universidad Autónoma de Madrid, 28049 Cantoblanco, Madrid Spain*

Room temperature photoluminescence and optical absorption spectra have been measured in wurtzite $In_xAl_yGa_{1-x-y}N$ (x ~ 0.06, 0.02 < y < 0.27) layers grown by molecular beam epitaxy. Photoluminescence spectra show both an enhancement of the integrated intensity and an increasing Stokes shift with the Al content. Both effects arise from an Al-enhanced exciton localization revealed by the S- and W-shaped temperature dependences of the photoluminescence emission energy and bandwidth respectively. Present results point to these materials as a promising choice for the active region in efficient light emitters. An In-related bowing parameter of 1.6 eV was derived from optical absorption data.

---

[a)] Electronic mail: sfernandez@die.upm.es

Efficient visible Light Emitting Diodes (LEDs) and Laser Diodes (LDs) based on InGaN alloys are commercially available despite a very high dislocation density due to heteroepitaxial growth. The high efficiency of InGaN-based LEDs and LDs is generally attributed to a strong In-induced exciton localization that prevents carriers to reach non-radiative recombination centers (NRCs).[1,2]

State of the art (Al)GaN ultraviolet (UV) LEDs show a much lower efficiency than InGaN-LEDs due to difficulties to achieve p-type conductivity with increasing Al content and to the lack of efficient exciton localization, that make carriers very sensitive to NRCs. Instead of the (Al)GaN system, quaternary InAlGaN alloys (QNs) have been proposed to improve UV-LEDs.[3] QNs provide a separate control of the bandgap energy and the lattice parameter, thus allowing a reduction of strain-related defects[4] and the built-in electric field in quantum well (QW) heterostructures grown along the $c$-axis.[5] In addition, In incorporation into AlGaN may induce exciton localization leading to an increasing efficiency (localization energy) for higher In concentrations.[2,6-8] However, QNs are not widely used yet due to difficulties to grow high quality alloys and the lack of knowledge on their properties.

This work reports on the optical properties of QNs with low In content (~ 6 %) as a function of the AlN mole fraction (2-27 %). Photoluminescence (PL) and optical absorption were studied to determine the dependence of the bandgap and the localization degree with the alloy composition. The observed room temperature (RT) PL enhancement with the Al content is discussed in terms of Al-enhanced exciton localization.

Wurtzite QN layers were grown by rf plasma-assisted molecular beam epitaxy (PA-MBE) on (0001) GaN templates (3.6 μm thick) grown by MOVPE on sapphire with a threading dislocation density (TDD) of ~ 1 x $10^8$ cm$^{-2}$ (Lumilog). Prior to the QN growth, a 100 nm thick GaN buffer layer was grown at 730 ºC to obtain a smooth and flat surface. Four QN samples (~ 300 nm thick), A, B, C and D, with increasing Al contents were grown at 600 ºC under intermediate metal-rich conditions to ensure a two-dimensional growth.[7,9] The AlN mole fraction was varied among samples keeping constant both the In and N fluxes, while the Ga flux was partially replaced by Al, being the III/V ratio the same for all samples.

Alloy compositions were determined by Rutherford backscattering spectrometry (RBS) using a 2 MeV He$^+$ ion beam. The spectra were acquired in random geometry and with the detector located at a scattering angle of 170º. High resolution x-ray diffraction (HRXRD) measurements were done to assess both the crystal quality and the strain state of the QN layers once their composition was known from RBS data considering the validity of Vegard's law for the lattice parameter. Atomic force microscopy (AFM) measurements were carried out to study the surface morphology and to assess the TDD. PL was excited with the 325 nm line of a He-Cd laser with 1.5 mW power. Absorbance data were derived from transmittance and reflectance spectra obtained with a Jasco V-650 spectrophotometer.

Figure 1 shows the measured and simulated RBS spectra for sample C. The individual contributions to the spectrum from each element are also shown, heavier elements detected at higher energies. Note that, the low signal from lighter elements such as Al and N (not shown in this energy scale for clarity purposes) with respect to the background, limits the sensitivity for such elements. This limitation comes from the higher cross-section of heavier elements such as In and Ga. For this reason, the composition is derived assuming an $In_xAl_yGa_{1-x-y}N$/GaN structure, hence, the Al content in the QN layer is calculated by the deficiency of In and Ga in a stoichiometric nitride. The simulation employs a homogenous In incorporation along the growth direction, which provides a reliable fitting of the experimental data. The alloy compositions for each sample are summarized in Table I. Notice that, though the In flux was nominally kept constant for all samples, its actual incorporation slightly decreases with the Al content. Similar results were reported by Monroy *et al.*[7,9] that claimed In segregation enhancement caused by the higher energy of Al-N bonds compared to those of Ga-N and, specially, of In-N.

The values of the full width at half maximum (FWHM) of symmetric (0002) $\omega$ and (0002) $\theta$-$2\theta$ scans are collected in Table I. The FWHM values of symmetric (0002) $\omega$, which are sensitive to both the grain size and the column tilt,[7] were similar to that of the GaN template indicating that the mosaicity of the QN layers is comparable with that of the GaN substrate. However, the FWHM values of the (0002) $\theta$-$2\theta$ were higher than that of the GaN template which points out a certain alloy disorder induced by the incorporation of In and Al into the layer.[7] The AFM measurements carried out in 5 x 5 $\mu m^2$ areas showed smooth surfaces with average roughness (rms) in the 0.6-1.1 nm range and a TDD comparable with that of the GaN template (Table I).

Figure 2(a) shows RT-PL spectra from the QNs and the GaN template (reference) revealing a single and broad band edge emission and the signature from the GaN template. The QNs PL emission blue-shifts from 3.14 (sample A) to 3.37 eV (sample D) with the Al content due to the bandgap increase, while the FWHM increases from 80 (sample A) to 140 meV (sample D), suggesting an enhancement of alloy fluctuations. The integrated PL intensity increases dramatically with the Al content (more than 2 orders of magnitude from sample A to D), being for sample D even three times higher than that of the GaN template.

Figure 2(b) shows RT absorbance spectra from QNs and the GaN template (reference). The absorption band edge is clearly seen in sample A, but it is partially overlapped with that of GaN in sample B. For samples C and D, the observed band edge is basically that of the GaN template, revealing a QN bandgap at least as wide as that of GaN. The unambiguous observation of the absorption band edge in sample A allows an estimation of its band gap by fitting the absorption spectrum, derived from transmittance (T) and reflectance (R) data [$\alpha(E) \alpha \ln(T/1-R)$], to the sigmoidal formula:[10]

$$\alpha(E) = \frac{\alpha_0}{1+\exp\left(\frac{E_G - E}{\Delta E}\right)} \quad (1)$$

where $E_G$ and $\Delta E$ represent the "effective band gap", and the absorption edge broadening, respectively. The QN band gap empirical dependence with composition and residual strain along the *c*-axis, $\epsilon_{zz}$, is given by:

$$E_G = xE_G^{InN} + yE_G^{AlN} + (1-x-y)E_G^{GaN} - b_{In}x(1-x) - b_{Al}y(1-y) + \frac{dE_G}{d\epsilon_{zz}}\epsilon_{zz} \quad (2)$$

where *x* and *y* are the In and Al mole fractions, $E_G^{InN}$ = 0.76 eV, $E_G^{AlN}$ = 6.16 eV, $E_G^{GaN}$ = 3.44 eV are the band gap energies of InN, AlN and GaN at RT[11], and $b_{Al}$ = 0.7 eV[11] and $b_{In}$ are the Al- and In-related bowing parameters respectively. Once known the band gap, the alloy composition, and $\epsilon_{zz}$, (assessed by HRXRD from a symmetric (0002) reflection) a value of $b_{In}$ = 1.6 ± 0.8 eV was derived assuming $dE_G/d\epsilon_{zz}$ = 15.4 eV as in GaN[12] due to the lack of experimental data for QNs. This value is much lower than the 3-17 eV already published in Ref. 13, but close within the experimental error to the 2.5 eV published in Ref. 7 and in good agreement with the 1.4 eV reported on InGaN.[11]

The comparison of PL and absorbance spectra for samples A and B shows a Stokes shift of ≈ 40 and ≈ 120 meV respectively. Comparable values were reported on both QNs[6,7] and InGaN[10,14] layers, explained in terms of exciton localization in potential fluctuations due to alloy inhomogenties. Note that for samples C and D it was not possible to estimate the Stokes shift because the absorption band edges correspond to the GaN template. However, for sample C a Stokes shift higher than 150 meV is expected considering that the QN bandgap must be at least as wide as that of GaN. Therefore, these results reveal an increasing Stokes shift for higher AlN mole fractions.

Figure 3 shows the PL emission energy and bandwidth temperature evolutions for each sample. In sample A (lowest Al content) the PL peak energy and the FWHM follow, above 30 K, an S- and W-shaped behavior respectively, which provide evidence of exciton localization in potential minima caused by alloy fluctuations.[6,8,14] Below 30 K the PL peak showed an anomalous blue-shift that was tentatively attributed (see Ref. 15) to exciton thermalization into shallower potential minima than those responsible of the S-shape. The same temperature dependence was observed in sample B but shifted towards higher temperatures indicating a stronger localization degree.[8] In samples with higher Al contents (C and D) the S- and W-shapes were not fully observed because localization was stronger and excitons remained partially localized even at RT. Thus, these results point to an exciton localization enhancement for higher AlN mole fractions despite the lower In incorporation into the layers. It is generally accepted that exciton localization in In containing III-nitride alloys takes place at "In-rich regions" within the In(Al)GaN matrix.[1-3,10,15] Therefore, the reported Al-enhanced exciton localization can be understood as a direct effect of the increasing well depth around the In-rich regions due to the higher average band gap of the InAlGaN matrix. In addition, the observed stronger localization degree for higher Al contents provides an explanation to the increasing Stokes shift[10] and to the RT-PL enhancement (carriers are less sensitive to NRCs[1,2]).

In summary, the optical properties of QN layers with low In content (~ 6%) were studied by PL and optical absorption as a function of the AlN mole fraction (2 to 27%).

The observed RT-PL integrated intensity enhancement and increasing Stokes shift with the Al content are explained by an Al-enhanced exciton localization. From RT optical absorption data an In-related bowing parameter of 1.6 eV was derived. These results suggest that the efficiency of visible and UV-LEDs and LDs may be improved taking advantage of the Al-enhanced exciton localization in QN layers.

## Acknowledgments

Thanks are due to Prof. K.H. Ploog for fruitful discussions. This work was partially supported by research grants from the Spanish Ministry of Education (MAT2004-2875, NAN04/09109/C04/2, Consolider-CSD 2006-19, and FPU program); and the Community of Madrid (GR/MAT/0042/2004 and S-0505/ESP-0200).

**Table Caption**

Table I. Alloy composition (from RBS data), FWHM values of $\omega$ and $\theta$-$2\theta$ x-ray scans for symmetric (0002) reflection, and average surface roughness (rms) and TDD assessed by AFM for InAlGaN layers, A, B, C and D, and GaN template.

| Sample | Al mole fraction | In mole fraction | (0002) $\Delta\omega$ [arcsec] | (0002) $\Delta\theta$-$2\theta$ [arcsec] | RMS [nm] | TDD [x $10^8$ cm$^{-2}$] |
|---|---|---|---|---|---|---|
| A | 0.02 | 0.08 | 325 | 165 | 1.1 | 2.1 |
| B | 0.13 | 0.06 | 336 | 103 | 0.8 | 1.9 |
| C | 0.20 | 0.06 | 335 | 170 | 0.8 | 1.6 |
| D | 0.27 | 0.05 | 334 | 147 | 0.7 | 1.7 |
| GaN Template | | | 322 | 90 | 0.4 | 1.0 |

**Figure captions**

Figure 1    Measured and simulated (solid line) RBS spectra for sample C. Dashed lines represent the individual contributions to the simulated spectrum from each element. The surface channels are also marked for the corresponding element.

Figure 2    (a) RT photoluminescence, and (b) absorbance spectra for InAlGaN layers, A, B, C and D, and GaN template, T.

Figure 3    PL emission energy (solid squares) and FWHM (open squares) as a function of the temperature for InAlGaN layers, A, B, C, and D. Dotted lines are guides to the eye.

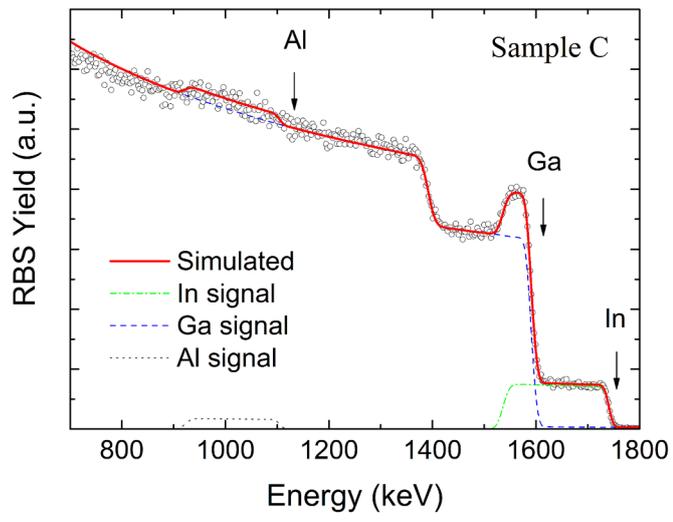

Figure 1

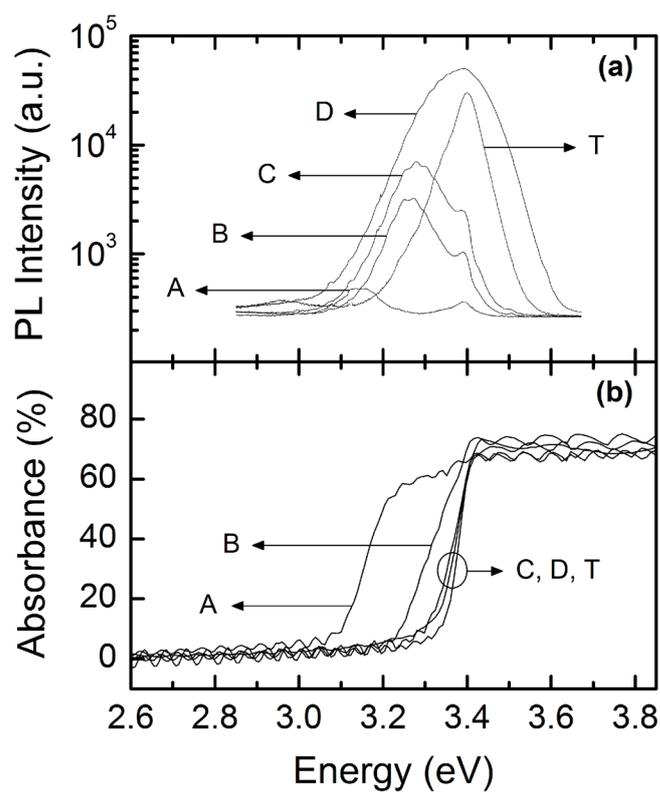

Figure 2

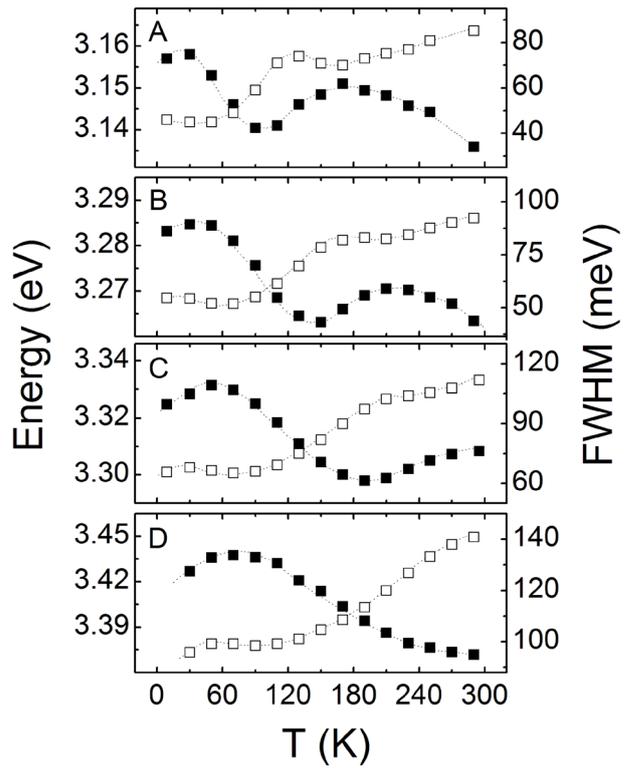

Figure 3